\documentclass[10pt,a4paper]{article}

\usepackage{gastex}

\usepackage{amsmath,amssymb,amscd,latexsym}
\usepackage{stmaryrd}
\usepackage{graphics, psfig,graphicx}
\usepackage{eufrak}
\usepackage{multind}

\usepackage[all,dvips]{xy}

\usepackage{theorem}

\numberwithin{equation}{section}
\newcommand\refeq[1]{(\ref{#1})}

\newtheorem{def.notation}{D\'efinition--Notations}[section]
\newtheorem{defff}{Definition}[section]

\newtheorem{prop}{Proposition}[section]
\newtheorem{prop.def}{Proposition--D\'efinition}[section]

\newtheorem{propriete.def}{Propri\'et\'e--D\'efinition}[section]
\newtheorem{property}{Property}[section]

\newenvironment{proof}{\noindent{\textsf{\underline{Proof}: }}}{$\blacksquare$}
\newtheorem{lemma}{Lemma}[section]
\newtheorem{theorem}{Theorem}[section]
\newtheorem{theorem*}{Theorem}

\theoremstyle{break}
{\theorembodyfont{\rmfamily}}
{\theorembodyfont{\rmfamily}}
{\theorembodyfont{\rmfamily}\newtheorem{example}{Example}[section]}
{\theorembodyfont{\rmfamily}}
{\theorembodyfont{\rmfamily}}
{\theorembodyfont{\rmfamily}}
{\theorembodyfont{\rmfamily}\newtheorem{NTO}{Notation}[section]}
{\theorembodyfont{\rmfamily}\newtheorem{remark}{Remark}[section]}
{\theorembodyfont{\rmfamily}\newtheorem{remarks}{Remarks}[section]}

\theoremstyle{plain}
\newtheorem{remark.num}{\sous{Remark}: }[section]

\newcommand\ie{i.e. }
\newcommand\eg{e.g. }

\newcommand\N{\mathbb{N}}
\newcommand\R{\mathbb{R}}

\newcommand\D{\partial}

\newcommand\dd{\text{d}}

\newcommand\opa{\textbf{\large{a}}}

\newcommand\NO[1]{\ensuremath{\Arrowvert #1 \Arrowvert}}

\newcommand\vect[1]{\overrightarrow{#1}}
\newcommand\sous[1]{\underline{#1}}

\newcommand\dsurd[2]{\frac{\partial #1}{\partial #2} }
\newcommand\chapo[1]{\widehat{#1} }
\newcommand\mc[1]{\ensuremath{\mathcal{#1}} }

\newcommand\Min{\mc{X}_{0}}

\newcommand\lang{\left\langle }
\newcommand\rang{\right\rangle }

\newcommand\intit{>\!\!\!\!\!\!\!\int}

\newcommand\Avide{\circ}

\newcommand\tree{\mathbb{T}}

\title{Perturbative classical and quantum field theory}
\author{Dikanaina HARRIVEL\footnote{LAREMA, UMR 6093, Universit\'e d'Angers, France.  
\textbf{dika@tonton.univ-angers.fr}}}

\begin{document}
\maketitle

\begin{abstract}
In a first part we study the $\phi^{p+1}$--field theory from the classical point of view. Using Butcher series we compute 
explicitly the perturbative expansion of the solutions and we prove that this expansion converges if the coupling constant is 
small enough. Then we show that we can formally recover the Heisenberg's interacting quantum field directly from this 
expansion. In other words we write the Heisenberg interacting field as a Butcher series. 
\end{abstract}

\vspace{ 1cm} 

\renewcommand\Min{\R^{1,d}}

\renewcommand\thetheorem{}
\renewcommand\theequation{\arabic{equation}}
\renewcommand\theremark{}

\section*{Introduction}

Butcher series are sums indexed by planar trees introduced by J. C. Butcher \cite{Butcher} in order to study and classify
\cite{Butcher.complements} Runge Kutta methods in numerical analysis. But they appear naturally in a large class of problem 
(see \eg \cite{dika.controle}, \cite{Brouder.BIT}) including quantum field theory (see \eg the paper of Ch. Brouder and A. Frabetti 
\cite{Brouder.arbre2}, \cite{ARBRE}, \cite{ARBRE2} about Q.E.D). Moreover Ch. Brouder noticed that the structure which underlies 
the original Butcher's calculations \cite{Butcher} is exactly the Hopf algebra of rooted trees defined by D. Kreimer in his paper 
about renormalization \cite{Kreimer} and he shows that finally renormalization can be seen as a Runge--Kutta method. 
Hence it seems that the work of J.C. Butcher and perturbative quantum field theory are closely related. In this paper  
we show that formally the Heisenberg's interacting quantum field can be written as a Butcher series. Hence we link the perturbative 
classical field theory with the perturbative quantum field theory. \\

Butcher series provide a precise description of the solutions of a non linear problem. In this paper we focus on 
$\phi^{p+1}$--theory but our results can be generalized to any theory. First, we explain how Butcher series give the explicit 
perturbative expansion of the classical field and we prove that this expansion converges if the coupling constant is small 
enough. Then, we formally quantize the expansion and we show that we recover the Heisenberg picture of the interacting quantum 
field. 

Let $p$ be a non negative integer such that $p\ge 2$. Then consider the equation over scalar fields 
$\varphi:\R^n\longrightarrow\R$ 
\begin{equation}\label{KGp.intro}\tag{K--Gp}
(\Box+m^2)\varphi+\lambda\varphi^p=0
\end{equation}
where $m>0$ is the mass, $\lambda$ is a real parameter (the coupling constant) and $\Box$ denotes the operator 
$\frac{\D}{\D (x^0)^2}-\sum_j\frac{\D}{\D (x^j)^2}$. We show that the solution of \refeq{KGp.intro} writes as a power 
series indexed by planar trees. \\

Planar trees are rooted trees drawn into the plane with the root on the ground. The external vertices are called \emph{leaves} 
and the other are called \emph{internal vertices}. We denote by $\vert b\vert$ the number of internal vertices of a planar tree 
$b$. There is a unique planar tree with no internal vertices: this is the planar tree reduced to a root, let's denote by $\circ$ 
this trees. We say that a planar tree is a \emph{$p$--tree} if and only if each internal vertex has exactly $p$ childrens. 
Let us denote by $\tree(p)$ the set of $p$--trees. Finally given $(b_1,\ldots,b_p)$ a $p$--uplet of $p$--trees, we can define 
another $p$--tree $B_+(b_1\ldots b_p)$ by linking a new root to the roots of $b_1$, \dots, $b_p$. \\

Given the Cauchy datas $(\varphi^0,\varphi^1)\in H^{q+1}(\R^d)\times H^q(\R^d)$, we show (theorem \ref{th.classique}) that 
the solution $\varphi\in\mc{C}^1([0,T],H^q(\R^d))\cap\mc{C}^2([0,T],H^{q-1}(\R^d))$ of \refeq{KGp.intro} such that 
$\varphi(0,\bullet)=\varphi^0$ and $\dsurd{\varphi}{t}(0,\bullet)=\varphi^1$ is given by the power series  
\begin{equation}\label{Butcher.intro}\tag{\mc{B}}
\varphi=\sum_{b\in\tree(p)}\lambda^{\vert b\vert}\phi(b)
\end{equation} 
where the coefficients $(\phi(b))_{b\in\tree(p)}$ are such that 
$$
\left\{
\begin{array}{l}
\phi(\circ)\text{ is the solution of }(\Box+m^2)\phi(\circ)=0 \text{ with Cauchy datas }(\varphi^0,\varphi^1)\\
\phi(B_+(b_1\ldots b_p))\text{ is the  solution of }(\Box+m^2)\varphi=-\phi(b_1)\cdots\phi(b_p)\text{ with zero Cauchy datas}
\end{array}
\right.
$$
The power series \refeq{Butcher.intro} converges in the $\mc{C}^1([0,T],H^q(\R^d))\cap\mc{C}^2([0,T],H^{q-1}(\R^d))$ topology if 
$\lambda$ is small enough, it is called the \emph{Butcher series}. 

\begin{remark}
The function $\phi(\circ)$ is the (classical) free field corresponding to the Cauchy datas $(\varphi^0,\varphi^1)$ at time $t=0$. 
We can reformulate the definition of $\phi(B_+(b_1\ldots b_p))$ by 
$$
\phi(B_+(b_1\ldots b_p))(x):=
-\int_{P^+}\dd y G_{ret}(x-y)\phi(b_1)(y)\cdots\phi(b_p)(y)
$$
where $G_{ret}(z)$ denotes the retarded Green function of the linear Klein--Gordon operator $(\Box+m^2)$ and where $P^+$ is the 
half space $P^+:=\{(t,\vect{x})\in\R\times\R^d\ \vert\ t>0\}$. \\
\end{remark}

Using the last remark we can formally quantize the Butcher series. We use the notation of the book 
of E. Peskin and Daniel V. Schroeder \cite{Peskin.Schroeder} p.83. \ie we consider the operator $\phi_I(x)$ acting on Fock space 
defined by 
$$
\phi_I(t,\vect{x}):=\frac{1}{(2\pi)^{d}}\int_{\R^{d}}\frac{\dd k}{\sqrt{2\omega_k}}
\left.\left(\opa_k e^{-ik\cdot x}+\opa_k^\dagger e^{i k\cdot x}\right)\right\vert_{x^0=t-t_0}
$$
Here $k\cdot x$ denotes the quantity $k\cdot x:=\omega_k t - \lang k,\vect{x}\rang$ where $\lang k,\vect{x}\rang$ denotes the 
usual scalar product of $\R^{d}$ and $\omega_k$ the quantity $\omega_k:=(\vert k\vert^2+m^2)^{1/2}$. 
The operator $\phi_I(x)$ is the free field on space--time. \\

Then by analogy we define formally the family of operator $(\chapo{\phi}(b))_{b\in\tree(p)}$ by 
$$
\left\{
\begin{array}{l}
\displaystyle{\chapo{\phi}(\circ)(x):=\phi_I(x)}\\
\displaystyle{\chapo{\phi}(B_+(b_1\ldots b_p))(x):=-\int_{P^+}\dd^{d+1} y G_{ret}(x-y)\chapo{\phi}(b_1)(y)\cdots\chapo{\phi}(b_p)(y)}
\end{array}
\right.
$$
Note that this definition is formal since operator product is ill defined on the diagonal. \\

Using the commutation rules of the $\phi_I(x)$'s, we get the following theorem: 
\begin{theorem}\hspace{-0.2cm}\textbf{\ref{Lien.Butcher-pQFT}}\sl{
for all $t>t_0$ and $\vect{x}\in\R^{d}$ we have formally 
$$
\chapo{\varphi}(t,\vect{x}):=\sum_{b\in \tree(p)}\lambda^{\vert b\vert} \chapo{\phi}(b)(x)=U^\dagger(t)\ \phi_I(t,\vect{x})\ U(t)
$$
where $U(t)$ denotes the operator 
$$
U(t):=\sum_{\alpha\ge 0}\left(-i\lambda\right)^\alpha 
\int_{t_O}^t\dd\tau_1\int_{t_0}^{\tau_1}\dd\tau_2\cdots\int_{t_0}^{\tau_{\alpha-1}}\dd\tau_\alpha 
H_I(\tau_1)H_I(\tau_2)\cdots H_I(\tau_\alpha)
$$
and $H_I(\tau)$ is given by 
$$
H_I(\tau):=\frac{1}{p+1}\int_{\R^{d}}\dd\vect{y}\phi_I^{p+1}(\tau,\vect{y})
$$ 
}
\end{theorem}
Using the time ordered product, the operator $U$ is given by the well known formula 
(see \cite{Peskin.Schroeder} p.85): 
$$
U(t)=T\left\{\exp\left(-i\lambda\int_{t_0}^t\dd\tau H_I(\tau)\right)\right\}
$$
Hence we recover the Heisenberg formulation of interacting quantum field as a quantization of Butcher series. 
From another point of view we prove that formally the interacting quantum field given by the Heisenberg identity can 
be expressed as a Butcher series. 

\renewcommand\thetheorem{\thesection.\arabic{theorem}}
\renewcommand\theequation{\thesection.\arabic{equation}}
\renewcommand\theremark{\thesection.\arabic{remark}}

\section{Planar $p$--trees}\label{p.arbres}
Let us introduce some definitions concerning planar trees. 

\begin{defff}\label{def.arbres.plans}\sl{
A \emph{planar tree} is an oriented connected finite graph without loop together with an embedding into the plane; we suppose  
that the graph has a particular node that no edge points to; this node is called the \emph{ root }Êof the tree. The 
root is drawn at the bottom of the tree. 
}
\end{defff}
\begin{remark}\label{remarque.rooted}
The set of planar trees differs from the set of \emph{rooted trees} used by D. Kreimer and A. Connes in their work about 
renormalization \cite{Kreimer}, \cite{Connes.Kreimer}. For instance the following planar trees are different 
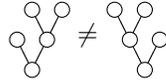
\begin{center}
\begin{picture}(0,6)(10,-2)
\gasset{Nw=2.0,Nh=2.0,Nmr=1.0,AHangle=0.0,AHLength=1.6}

\node(n1)(4,4){}
\node(n2)(8,4){}
\node(n3)(2,0){}
\node(n4)(6,0){}
\node(n5)(4,-4){}
\drawedge(n1,n4){}
\drawedge(n2,n4){}
\drawedge(n3,n5){}
\drawedge(n4,n5){}

\put(10,-0.3){$\neq$}

\node(n11)(15,4){}
\node(n12)(19,4){}
\node(n13)(17,0){}
\node(n14)(21,0){}
\node(n15)(19,-4){}
\drawedge(n11,n13){}
\drawedge(n12,n13){}
\drawedge(n13,n15){}
\drawedge(n14,n15){}

\end{picture}
\end{center}
although they represent the same rooted tree. All the results of this paper can be expressed using rooted trees instead of planar 
trees: we just have to adapt the definitions and add a symmetry factor in front of each term of the perturbative expansions. 
We prefer to work with planar trees because they permit to avoid these symmetry factors and to get simpler formulaes. 
\end{remark}

\begin{NTO}\label{notation.arbres.plans}\sl{
Let $b$ be a planar tree
\begin{enumerate}
\item The external vertices of $b$ are called \emph{leaves} and the other vertices \emph{internal vertices}, $\vert b\vert$ 
denotes the number of internal vertices of $b$.
\item We denote by $\circ$ the planar tree without internal vertex. 
\item A planar tree is a \emph{$p$--tree} if each of its internal vertices has \emph{exactly} $p$ childrens. We denote by 
$\tree(p)$ the set of $p$--trees. 
\end{enumerate}
}
\end{NTO}
\begin{example}\label{exemple.nb}\sl{
The planar trees of remark \ref{remarque.rooted} are $2$--trees and they satisfy $\vert b\vert=2$. 
}
\end{example}

\begin{defff}\label{def.B_+}\sl{
Let $(b_1,\ldots,b_p)\in \tree(p)^p$, then we denote by $B_+(b_1,\ldots,b_p)$ the $p$-tree obtained by 
connecting to a new root the roots of $b_1$ and $b_2$ and \dots and $b_p$. 
\begin{center}
\begin{picture}(93,13)(0,-22)
\gasset{linewidth=0.3,Nw=8.0,Nh=8.0,Nmr=0.0}
\node(n0)(36.11,-12.12){$b_1$}
\node(n1)(48.11,-12.12){$b_2$}
\node(n3)(84.11,-12.12){$b_p$}
\node[linewidth=0.3,Nw=2.0,Nh=2.0,Nmr=1.0](n4)(60.11,-26){}
\put(60,-12){$\cdots\cdots$}
\gasset{AHangle=0.0}
\drawedge(n0,n4){}
\drawedge(n1,n4){}
\drawedge(n3,n4){}
\put(0,-18.0){$B_+(b_1,\ldots,b_p)=$}
\end{picture}
\end{center}
}
\end{defff}

\begin{property}\label{prop.recusrsion.arbre}\sl{
Let $b\in\tree$ be such that $b\neq\circ$ then there exists a unique $p$--uplet $(b_1,\ldots,b_p)\in \tree(p)^p$ such that 
$b=B_+(b_1,\ldots,b_p)$. 
}
\end{property}

\section{Perturbative classical field theory: Butcher series}\label{classique}
Let $q$ be an integer such that $q>d/2$. Then it is well known (see \eg \cite{ADAMS}) that the Sobolev space $H^q(\R^{d})$ 
is an algebra, more precisely we have 
\begin{prop}\label{algebre}\sl{
Let $f$ and $g$ belong to $H^q(\R^{d})$ then the product $fg$ still belongs to $H^q(\R^{d})$ and 
there exists a constant $c_q>0$ which depends only on $q$ and $d$ such that 
$$
\NO{fg}_{H^q(\R^{d})}\le c_q\NO{f}_{H^q(\R^{d})}\NO{g}_{H^q(\R^{d})}
$$
}
\end{prop}
From now on, for all $q\in\N$, $H^q$ denotes the Sobolev space $H^q(\R^{d})$. \\

Let $T>0$ and $(\varphi^0,\varphi^1)\in H^{q+1}\times H^q$. Consider the following problem 
\begin{equation}\label{pb.classique}
\left\{
\begin{array}{l}
\displaystyle{\varphi\in\mc{C}^1([0,T],H^q)\cap\mc{C}^2([0,T],H^{q-1})}\\
\displaystyle{(\Box+m^2)\varphi+\lambda\varphi^p=0\text{ in }H^{q-1}}\\
\displaystyle{\varphi(0,\bullet)=\varphi^0\ ;\ \dsurd{\varphi}{t}(0,\bullet)=\varphi^1}
\end{array}
\right.
\end{equation}
We define recursively the family $(\phi(b))_{b\in\tree(p)}$ by  setting 
\begin{equation}\label{initial.classique}
\chapo{\phi(\circ)}(t,\vect{k}):=\frac{\sin(\omega_k t)}{\omega_k}\chapo{\varphi^1}(\vect{k})
+cos(\omega_k t)\chapo{\varphi^0}(\vect{k}) 
\end{equation}
where for all $\psi\in\mc{C}^0([0,T],\R^{d})$, $\chapo{\psi}(t,\vect{k})$ denotes the spatial Fourier transform of $\psi$. 
For all $(b_1,\ldots,b_p)\in\tree(p)^p$ we set for all $x\in\R^n$ 
\begin{equation}\label{recursion.classique}
\phi(B_+(b_1,\ldots,b_p))(x):=
-\int_{P_+}G_{ret}(x-y)\phi(b_1)(y)\cdots\phi(b_p)(y)
\end{equation}
where $P_+:=\{(t,\vect{x})\in\R^n\vert t>0\}$ and $G_{ret}$ denotes the retarded Green function of the Klein--Gordon operator \ie 
\begin{equation}\label{G.ret}
G_{ret}(z):=\frac{1}{(2\pi)^{d}}\theta(z^0)\int_{\R^{d}}\dd^{d} k\frac{\sin(z^0\omega_{k})}{\omega_{k}}e^{ik.\vect{z}}
\dd^n y
\end{equation} 
$\theta$ denotes the heavyside function ($\theta(t)=0$ if $t<0$ and $1$ otherwise). 
\begin{remark}
We can easily check that $\phi(\circ)$ is the solution of problem \refeq{pb.classique} with $\lambda=0$ \ie 
\begin{gather*}
(\Box+m^2)\phi(\circ)=0\\
\phi(\circ)(0,\bullet)=\varphi^0\text{ ; }\dsurd{\phi(\circ)}{t}(0,\bullet)=\varphi^1
\end{gather*}
So we can consider $\phi(\circ)$ as "\emph{the free field corresponding to the interacting field $\varphi$ at time $t=0$}". 

In other hand $\phi(B_+(b_1\ldots b_p))$ satisfies 
$$
(\Box+m^2)\phi(B_+(b_1\ldots b_p))=-\phi(b_1)\cdots\phi(b_p)
$$
with zero Cauchy data on the hypersurface $t=0$. 
\end{remark}

Then we have the following result 
\begin{theorem}\label{th.classique}\sl{
For all $T>0$ and $(\varphi^0,\varphi^1)\in H^{q+1}\times H^q$ the family $(\phi(b))_{b\in\tree(p)}$ is well defined by 
\refeq{initial.classique} and \refeq{recursion.classique} and $\forall b\in\tree(p)$, 
$\phi(b)$ belongs to $\mc{C}^1([0,T],H^q)\cap\mc{C}^2([0,T],H^{q-1})$. Moreover there exists a constant $C>0$ which depends only 
on $m$, $d$, $p$ and $q$ such that if 
\begin{equation}\label{condition.convergence.theo}
C\vert \lambda (1+MT)\vert \left(\NO{\varphi^0}+\NO{\varphi^1}\right)^{p-1}<1
\end{equation}
(here $M$ denotes the constant $M:=\max(m,1/m)$) then the power series 
$$
\varphi=\sum_{b\in\tree(p)}\lambda^{\vert b\vert}\phi(b)
$$ 
converges in the $\mc{C}^1([0,T],H^q)\cap\mc{C}^2([0,T],H^{q-1})$ topology and the sum $\varphi$ is a solution of 
problem \refeq{pb.classique}. 
}
\end{theorem}
\begin{remarks}
We have studied the equation \refeq{pb.classique} but our approach can be extended for analytic non linearity. 
In this case, all planar trees are involved and condition \refeq{condition.convergence.theo} must be adapted 
(see \cite{dika.controle} for more details). 
\end{remarks}
\begin{proof}(of theorem \ref{th.classique})\\
The proof of theorem \refeq{th.classique} is very simple. Suppose that the power series 
$\varphi=\sum_{b\in\tree(p)}\lambda^{\vert b\vert} \phi(b)$ converges then a simple calculation show that we have 
$\varphi(0,\bullet)=\phi(\circ)(0,\bullet)+0=\varphi^0$ and 
$\dsurd{\varphi}{t}(0,\bullet)=\dsurd{\phi(\circ)}{t}(0,\bullet)+0=\varphi^1$ and for all $(b_1\ldots b_p)\in\tree(p)^p$
$$
(\Box+m^2)\phi(B_+(b_1,\ldots,b_p))=-\phi(b_1)\ldots\phi(b_p)
$$
But for all $b\in\tree(p)$, $b\neq\circ$ there exists a unique $p$--uplet $(b_1\ldots b_p)$ of $p$--tree such that 
$b=B_+(b_1\ldots b_p)$ hence we get 
\begin{align}
(\Box+m^2)\varphi
=&(\Box+m^2)\phi(\circ)+\sum_{(b_1,\ldots,b_p)\in\tree(p)^p}\lambda^{\vert B_+(b_1,\ldots,b_p)\vert}
(\Box+m^2)\phi(B_+(b_1,\ldots,b_p))\\
=&0-\lambda \sum_{(b_1,\ldots,b_p)\in\tree(p)^p} \lambda^{\vert b_1\vert+\cdots+\vert b_p\vert}\phi(b_1)\cdots\phi(b_p)
=-\lambda\varphi^p
\end{align}

Let us focus on the convergence of the power series. Let us show recursively that $\phi(b)$ belongs to 
$\mc{C}^1([0,T],H^q)\cap\mc{C}^2([0,T],H^{q-1})$ for all $b\in\tree(p)$ and 
\begin{equation}\label{dem.classique}
\NO{\phi(b)}\le \left(c_q^{p-1}(1+MT)\right)^{\vert b\vert} 
\left[M^2\left(\NO{\varphi^0}+\NO{\varphi^1}\right)\right]^{\vert b\vert(p-1)+1}
\end{equation}
where $M$ denotes the constant $M:=\max(m,1/m)\ge 1$. 

Let us show \refeq{dem.classique} for $b=\circ$. The function $\phi(\circ)$ is given by \refeq{initial.classique}. Then we can 
easily show that $\phi(\circ)$ belong to $\mc{C}^1([0,T],H^q)\cap\mc{C}^2([0,T],H^{q-1})$ and since 
$(m^2+\alpha^2)/(1+\alpha^2)\le M^2$ for all $\alpha\in\R$ we have 
\begin{gather*}
\NO{\phi(\circ)(t,\bullet)}_{H^q}\le M\NO{\varphi^1}_{H^q}+\NO{\varphi^0}_{H^q}
\le M^2\left(\NO{\varphi^0}+\NO{\varphi^1}\right)\\
\NO{\dsurd{\phi(\circ)}{t}(t,\bullet)}_{H^q}\le \NO{\varphi^1}_{H^q}+M\NO{\varphi^0}_{H^{q+1}}
\le M^2\left(\NO{\varphi^0}+\NO{\varphi^1}\right)\\
\NO{\frac{\D^2\phi(\circ)}{\D t^2}(t,\bullet)}_{H^{q-1}}\le M\NO{\varphi^1}_{H^q}+M^2\NO{\varphi^0}_{H^{q+1}}
\le M^2\left(\NO{\varphi^0}+\NO{\varphi^1}\right)
\end{gather*}
so $\NO{\phi(\circ)}:=\max_{t\in[0,T]}\left(\NO{\phi(\circ)(t,\bullet)}_{H^q};\NO{\dsurd{\phi(\circ)}{t}(t,\bullet)}_{H^q}; 
\NO{\frac{\D^2\phi(\circ)}{\D t^2}(t,\bullet)}_{H^{q-1}}\right)$ satisfies \refeq{dem.classique}.

Suppose that \refeq{dem.classique} is satisfied for all $b\in\tree(p)$, $\vert b\vert\le N$. Let $b\in\tree(p)$ such that 
$\vert b\vert=N+1\ge 1$ then $\exists (b_1\ldots b_p)\in\tree(p)^p$ such that $b=B_+(b_1\ldots b_p)$ and 
$\phi(b)$ is defined by \refeq{recursion.classique} hence, using proposition \ref{algebre} we get 
\begin{gather*}
\NO{\phi(b)(t,\bullet)}_{H^q}\le c_q^{p-1}MT\NO{\phi(b_1)}\cdots\NO{\phi(b_p)}
\le (1+MT)c_q^{p-1}\NO{\phi(b_1)}\cdots\NO{\phi(b_p)}\\
\NO{\dsurd{\phi(b)}{t}(t,\bullet)}_{H^q}\le c_q^{p-1}T\NO{\phi(b_1)}\cdots\NO{\phi(b_p)}
\le (1+MT)c_q^{p-1}\NO{\phi(b_1)}\cdots\NO{\phi(b_p)}\\
\NO{\frac{\D^2\phi(b)}{\D t^2}(t,\bullet)}_{H^{q-1}}\le c_q^{p-1}(1+MT)
\NO{\phi(b_1)}\cdots\NO{\phi(b_p)}
\end{gather*}
hence we have $\NO{\phi(b)}\le (1+MT)c_q^{p-1}\NO{\phi(b_1)}\cdots\NO{\phi(b_p)}$. Since \refeq{dem.classique} is 
satisfied for $b_1$, \dots, $b_p$ and using the fact that 
$\vert B_+(b_1\ldots b_p)\vert=\vert b_1\vert+\cdots+\vert b_p\vert+1$ we finally get \refeq{dem.classique} for $b$. \\

Finally it can be shown (see \eg \cite{SEDGEWICK}) that the number of $p$--trees $b$ such that $\vert b\vert=N$ is bounded 
by  $(p^{p}/(p-1)^{p-1})^N$ hence we see that if we have 
\begin{equation}\label{condition.convergence}
(1+MT)\vert\lambda\vert\left(\NO{\varphi^0}+\NO{\varphi^1}\right)^{p-1}<\frac{(p-1)^{p-1}}{p^{p}c_q^{p-1}M^{2(p-1)}}
\end{equation}
then the power series $\sum_{N} \vert \lambda\vert^N \NO{\phi(b)}$ converges which completes the proof. 
\end{proof}
\begin{remark}
Notice that the proof gives an explicit value for $C$. 
\end{remark}

\section{From Butcher series to perturbative quantum field theory}\label{perturbation.quantique}
We have seen that Butcher series provides an explicite and precise perturbative 
expansion of the \emph{solutions} of interacting Klein--Gordon equation 
\begin{equation}\label{ee}\tag{$E_p$}
(\Box+m^2)\varphi+\lambda\varphi^p=0
\end{equation}
So we get a perturbative \emph{classical} field theory. \\

On the other hand, physicists devollopped a perturbative quantum field theory 
(see \eg \cite{Peskin.Schroeder}, \cite{ITZAK}, \cite{RYDER}). 
The question we are interesting in is the following: is there a link between Butcher series and perturbative quantum field 
theory ? \\

The answer is yes, we will show that the Heisenberg picture of interacting quantum field can be 
written as a quantized Butcher series. \\

Let $t_0\in\R$ be a fixed time and $x=(t,\vect{x})\in\Min$. Then following M. Peskin et D. Schroeder 
\cite{Peskin.Schroeder} p. 83, the free field on space--time is the self--adjoint operator $\phi_I(x)$ acting on 
\emph{Fock space} (see \cite{Fock.rigoureux.1}, \cite{Fock.rigoureux.2}) defined by 
$$
\phi_I(x)=\phi_I(t,\vect{x}):=\frac{1}{(2\pi)^{d}}\int_{\R^{d}}\frac{\dd k}{\sqrt{2\omega_k}}
\left.\left(\opa_k e^{-ik\cdot x}+\opa_k^\dagger e^{i k\cdot x}\right)\right\vert_{x^0=t-t_0}
$$
Here $k\cdot x$ denotes the quantity $k\cdot x:=\omega_k t - \lang k,\vect{x}\rang$ where $\lang k,\vect{x}\rang$ denotes the 
usual scalar product on $\R^{d}$ and $\omega_k$ the quantity $\omega_k:=(\vert k\vert^2+m^2)^{1/2}$. The operator $\opa_k$ 
and $\opa_k^\dagger$ denote the usual \emph{creation} and \emph{annihilation} operator (see \eg \cite{Fock.rigoureux.2}). \\


We formally define the family $(\chapo{\phi}(b)(x))_{b\in \tree(p)}$ of operators by the following 
\begin{defff}\label{def.phi.quantique}\sl{
$\chapo{\phi}(\Avide)(x):=\phi_I(x)$ and for all $(b_1,\ldots,b_p)\in \tree(p)^p$ 
\begin{equation}\label{def.Butcher.quantique}
\chapo{\phi}(B_+(b_1,\ldots,b_p))(t,\vect{x}):=
-\int_{t_0}^{t}\dd y^0\int_{\R^{d}}\dd\vect{y} G_{ret}(x-y)\chapo{\phi}(b_1)(y)\cdots\chapo{\phi}(b_p)(y)
\end{equation}
}
\end{defff}
Hence we can see the sum $\sum_{b\in\tree(p)} \lambda^{\vert b\vert}\chapo{\phi}(b)(x)$ as the formal quantization of the Butcher series 
$\sum_{b\in\tree(p)}\lambda^{\vert b\vert}\phi(b)$. 
\begin{remark}\label{remarque.clef}
\begin{itemize}
\item The definition \ref{def.phi.quantique} is formal because we do not care about the definition of the operator product 
which is well known to be ill defined. 

\item We have the following commutation relation between the operators $\phi_I(x)$ and $\phi_I(y)$ 
\begin{equation}\label{commutation.champslibre.espace-temps}
\Delta(x-y):=\left[\phi_I(x),\phi_I(y)\right]=\frac{1}{(2\pi)^{d}}\int_{\R^{d}} \frac{\dd k}{2\omega_k}
\left(e^{-ik\cdot (x-y)}-e^{ik\cdot (x-y)}\right)
\end{equation}
(see \cite{Peskin.Schroeder} p.28). Using the expression \refeq{G.ret} of the retarded Green function $G_{ret}$, we get 
\begin{equation}\label{clef.Butcher.quantique}
G_{ret}(z)=-i\theta(\tau)\Delta(z)
\end{equation}
This simple remark leads to the following theorem: 
\end{itemize}
\end{remark}
\begin{theorem}\label{Lien.Butcher-pQFT}\sl{
for all $t>t_0$ and $\vect{x}\in\R^{d}$ we have 
\begin{equation}\label{Heisenberg}
\chapo{\varphi}(t,\vect{x}):=\sum_{b\in \tree(p)}\lambda^{\vert b\vert} \chapo{\phi}(b)(x)=U^\dagger(t)\ \phi_I(t,\vect{x})\ U(t)
\end{equation}
where $U(t)$ denotes the operator 
\begin{equation}\label{U}
U(t):=\sum_{\alpha\ge 0}\left(i\lambda\right)^\alpha 
\int_{t_0}^t\dd\tau_1\int_{t_0}^{\tau_1}\dd\tau_2\cdots\int_{t_0}^{\tau_{\alpha-1}}\dd\tau_\alpha 
H_I(\tau_1)H_I(\tau_2)\cdots H_I(\tau_\alpha)
\end{equation}
and $H_I(\tau)$ is given by 
$$
H_I(\tau):=\frac{1}{p+1}\int_{\R^{d}}\dd\vect{y}\phi_I^{p+1}(\tau,\vect{y})
$$ 
}
\end{theorem}
\begin{remarks}
\begin{itemize}
\item 
Using the time ordered product (see \cite{Peskin.Schroeder} p.85 for a definition) the operator $U$ is given by the well known 
formula 
$$
U(t)=T\left\{\exp\left(i\lambda\int_{t_0}^t\dd\tau H_I(\tau)\right)\right\}
$$
\item 
The right hand side of \refeq{Heisenberg} is exactly the Heisenberg picture of interacting quantum field 
and this can be the beginning of the perturbative quantum field theory (see \eg \cite{Peskin.Schroeder} p.77-87). 
\end{itemize}
\end{remarks}

\begin{proof}(of theorem \ref{Lien.Butcher-pQFT})\\
Let us introduce a new notation which will be very helpful for manipulations of iterated integrals such as \refeq{U}. 
Let $r\in\N^*$ and $t_0\le t$. Then the symbol $\intit_{t_0}^t\dd y_1\cdots\dd y_r$ 
denotes the integrals 
$$
\intit_{t_0}^t\dd y_1\cdots\dd y_r:=\int_{t_0}^t\dd y^0_1\int_{t_0}^{\tau_1}\dd y^0_2\cdots\int_{t_0}^{\tau_{p-1}}\dd y^0_p
\int_{\R^{d}}\dd^{d}\vect{y_1}\cdots\int_{\R^{d}}\dd^{d}\vect{y_r}
$$
Hence using this notation, expression \refeq{U} becomes 
\begin{equation}\label{U.2}
U(t)=\sum_{\alpha\ge 0}\left(\frac{i\lambda}{p+1}\right)^\alpha \intit_{t_0}^t\dd y_1\cdots\dd y_\alpha 
\ \phi_I^{p+1}(y_1)\phi_I^{p+1}(y_2)\cdots \phi_I^{p+1}(y_\alpha)
\end{equation}
So since $\phi_I(x)$ is self--adjoint we get the following identity 
\begin{multline*}
U^\dagger(t)\phi_I(t,\vect{x})U(t)=
\sum_{m\ge 0}\left(\frac{i\lambda}{p+1}\right)^m
\sum_{\substack{(r,s)\in\N \\ r+s=m}}\\(-1)^r 
\intit_{t_0}^t\dd y_1\cdots\dd y_r\intit_{t_0}^t\dd z_1\cdots\dd z_s
\phi_I^{p+1}(y_r)\cdots \phi_I^{p+1}(y_1)\phi_I(x)\phi_I^{p+1}(z_1)\cdots \phi_I^{p+1}(z_s)
\end{multline*}

To complete the proof of theorem \ref{Lien.Butcher-pQFT}, it suffices to show that for all $m\in\N$ 
\begin{multline}\label{ce.qu.on.veux.Heisenberg}
\sum_{\substack{b\in \tree(p)\\ \vert b\vert =m}}
\chapo{\phi}(b)(x)=\left(\frac{i}{p+1}\right)^m
\sum_{\substack{(r,s)\in\N \\ r+s=m}}(-1)^r 
\intit_{t_0}^t\dd y_1\cdots\dd y_r\intit_{t_0}^t\dd z_1\cdots\dd z_s\\
\phi_I^{p+1}(y_r)\cdots \phi_I^{p+1}(y_1)\phi_I(x)\phi_I^{p+1}(z_1)\cdots \phi_I^{p+1}(z_s)
\end{multline}
Let us prove \refeq{ce.qu.on.veux.Heisenberg} by induction on $m\in\N$. \\

For $m=0$ identity \refeq{ce.qu.on.veux.Heisenberg} is obvious since the left hand side of \refeq{ce.qu.on.veux.Heisenberg} 
reduces to $\phi_I(x)$ and the only $p$--tree such that $\vert b\vert=0$ is given by $b=\circ$ therefore the 
right hand side equals to $\chapo{\phi}(\Avide)(x)=\phi_I(x)$.

Let $N\in\N$ and suppose that identity \refeq{ce.qu.on.veux.Heisenberg} is satisfied for all $m\le N$. Then set 
$$
\varphi_{N+1}(x):=\sum_{\substack{b\in \tree(p)\\\vert b\vert=N+1}}\chapo{\phi}(b)(x)
$$
Since $N+1\ge 1$, for all $b\in\tree(p)$ such that $\vert b\vert=N+1$ there is a unique $(b_1,\ldots,b_p)\in \tree(p)^p$ such 
that $b=B_+(b_1,\ldots,b_p)$. So the definition of $\chapo{\phi}(b)(x)$ together with remark \ref{remarque.clef} lead to 
$$
\varphi_{N+1}=\sum_{\substack{(b_1,\ldots,b_p)\in \tree(p)^p\\ \vert b_1\vert+\cdots+\vert b_p\vert=N}}
i\int_{t_0}^t\dd y^0\int_{\R^{d}}\dd\vect{y}\Delta(x-y)
\chapo{\phi}(b_1)(y)\cdots\chapo{\phi}(b_p)(y)
$$
which can be rewritten as 
$$
\varphi_{N+1}=\sum_{\substack{(q_1,\ldots,q_p)\in\N^p\\ q_1+\cdots+q_p=N}}
i\int_{t_0}^t\dd y^0\int_{\R^{d}}\dd\vect{y}\Delta(x-y)
\bigg(\sum_{\substack{b_1\in \tree(p)\\ \vert b_1\vert=q_1}}\chapo{\phi}(b_1)(y)\bigg)\cdots
\bigg(\sum_{\substack{b_1\in \tree(p)\\ \vert b_p\vert=q_p}}\chapo{\phi}(b_p)(y)\bigg).
$$
Then using \refeq{ce.qu.on.veux.Heisenberg} for $m\le N$, we see that $\varphi_{N+1}$ is given by the following expression  
\begin{multline}\label{somme.monstrueuse}
i \left(\frac{i}{p+1}\right)^N
\sum_{\substack{(q_1,\ldots,q_p)\in\N^p\\ q_1+\cdots+q_p=N}}
\sum_{\substack{(r_1,s_1)\in\N^2\\ r_1+s_1=q_1}}\cdots \sum_{\substack{(r_p,s_p)\in\N^2\\ r_p+s_p=q_p}}(-1)^{r_1+\cdots+r_p}\\
\int_{t_0}^t\dd y^0\int_{\R^{d}}\dd\vect{y}\Delta(x-y)
\intit_{t_0}^{y^0}\dd y^{(1)}_{1,r_1}\intit_{t_0}^{y^0}\dd z^{(1)}_{1,s_1}\cdots 
\intit_{t_0}^{y^0}\dd y^{(p)}_{1,r_p}\intit_{t_0}^{y^0}\dd z^{(p)}_{1,s_p} \mathbb{P}((y_j^{(k)})_{j,k},y)
\end{multline}
Here we must explain the notations:
\begin{itemize}
\item Let $\alpha$, $\beta$ be some integers such that $\alpha\le \beta$, then we denote by $y_{\alpha,\beta}$ the 
$(\beta-\alpha+1)$--uplet $y_{\alpha,\beta}=(y_\alpha,\ldots,y_\beta)\in(\R^n)^{\beta-\alpha+1}$. 
\item $\mathbb{P}((y_j^{(k)})_{j,k},y)$ denotes the product 
\begin{multline*}
\phi_I^{p+1}(y^{(1)}_{r_1})\cdots\phi_I^{p+1}(y^{(1)}_{1})\phi_I(y)
\left[\phi_I^{p+1}(z_1^{(1)})\cdots\phi_I^{p+1}(z_{s_1}^{(1)})
\phi_I^{p+1}(y_{r_2}^{(2)})\cdots\phi_I^{p+1}(y_{1}^{(2)})\right]\phi_I(y)\\
\left[\phi_I^{p+1}(z_1^{(2)})\cdots \phi_I^{p+1}(z_{s_2}^{(2)})\phi_I^{p+1}(z_{r_3}^{(3)})\cdots\phi_I^{p+1}(z_{1}^{(3)}) \right]
\phi_I(y)\cdots \phi_I(y) \phi_I^{p+1}(z_{1}^{(p)})\cdots\phi_I^{p+1}(z_{s_p}^{(p)})
\end{multline*}
\end{itemize}
Then we use the well known combinatorial lemma of quantum field theory (see \eg \cite{Peskin.Schroeder}) which can be proved 
directly using iterated integrals 
\begin{lemma}\label{U.unitaire}\sl{
For all $t\ge 0$ we have $U(t)U^\dagger(t)=Id$ \ie for all $m\in\N$ 
\begin{equation*}
\sum_{\substack{(r,s)\in\N \\ r+s=m}}(-1)^r\intit_{t_0}^t\dd y'_1\ldots\dd y'_r\intit_{t_O}^t \dd y_1\ldots\dd y_s 
\phi_I^{p+1}(y_1)\cdots \phi_I^{p+1}(y_s) \phi_I^{p+1}(y'_r)\cdots \phi_I^{p+1}(y'_1)=\delta_{0,m}
\end{equation*}
where $\delta_{0,m}=0$ if $m\neq 0$ and $1$ otherwise. 
}
\end{lemma}
Using lemma \ref{U.unitaire} we see that expression \refeq{somme.monstrueuse} reduces to 
\begin{multline}\label{somme.moins.monstrueuse}
\varphi_{N+1}=i \left(\frac{i}{p+1}\right)^N
\sum_{\substack{(r,s)\in\N^2\\ r+s=N}}(-1)^r
\int_{t_0}^t\dd y^0\int_{\R^{d}}\dd\vect{y}\Delta(x-y)\\
\intit_{t_0}^{y^0}\dd y_1\cdots\dd y_r\intit_{t_0}^{y^0}\dd z_1\cdots\dd z_s
\ \phi_I^{p+1}(y_{r})\cdots\phi_I^{p+1}(y_{1})\phi_I^p(y)\phi_I^{p+1}(z_{1})\cdots\phi_I^{p+1}(z_{s})
\end{multline}
But since $\Delta(x-y)=[\phi_I(x),\phi_I(y)]$ which commutes with $\phi_I(z)$ (it is a $c$--number), we can replace 
$\Delta(x-y)\phi_I(y)^p$ in \refeq{somme.moins.monstrueuse} by its expression using $\phi_I(x)$ and $\phi_I(y)$ \ie 
$$
\Delta(x-y)\phi_I(y)^p=\phi_I(x)\phi_I^{p+1}(y)-\phi_I(y)\phi_I(x)\phi_I^p(y).
$$
But we have $\phi_I(x)\phi_I^p(y)=\phi_I^p(y)\phi_I(x)+p\Delta(x-y)\phi_I^{p-1}(y)$ therefore the last identity leads to  
$$
\Delta(x-y)\phi_I(y)^p=\phi_I(x)\phi_I^{p+1}(y)-\phi_I^{p+1}(y)\phi_I(x)-p\Delta(x-y)\phi_I^p(y).
$$
Inserting this last expression in \refeq{somme.moins.monstrueuse} we finally get 
\begin{multline}\label{somme.de.nouveau.monstrueuse}
\varphi_{N+1}= \left(\frac{i}{p+1}\right)^{N+1}
\sum_{\substack{(r,s)\in\N^2\\ r+s=N}}(-1)^r
\int_{t_0}^t\dd y^0\int_{\R^{d}}\dd\vect{y}
\intit_{t_0}^{y^0}\dd y_1\cdots\dd y_r\intit_{t_0}^{y^0}\dd z_1\cdots\dd z_s\\
\left[\phi_I^{p+1}(y_{r})\cdots\phi_I^{p+1}(y_{1})\phi_I(x)\phi_I^{p+1}(y)\phi_I^{p+1}(z_{1})\cdots\phi_I^{p+1}(z_{s})\right.
\\-
\left.\phi_I^{p+1}(y_{r})\cdots\phi_I^{p+1}(y_{1})\phi_I^{p+1}(y)\phi_I(x)\phi_I^{p+1}(z_{1})\cdots\phi_I^{p+1}(z_{s})\right]
\end{multline}

Consider separately the terms of the right hand side of \refeq{somme.de.nouveau.monstrueuse}. The first term is given by 
(modulo a factor  $(i/(p+1))^{N+1}$)  
\begin{multline}\label{premier.terme.monstre}
\sum_{\substack{(r,s)\in\N^2\\ r+s=N}}(-1)^r
\int_{t_0}^t\dd y^0\int_{\R^{d}}\dd\vect{y}
\intit_{t_0}^{y^0}\dd y_1\cdots\dd y_r\intit_{t_0}^{y^0}\dd z_1\cdots\dd z_s\\
\phi_I^{p+1}(y_{r})\cdots\phi_I^{p+1}(y_{1})\phi_I(x)\phi_I^{p+1}(y)\phi_I^{p+1}(z_{1})\cdots\phi_I^{p+1}(z_{s})
\end{multline}
Notice that for all $y\in[0,t]\times\R^{d}$ and $a\ge 1$ we have 
$$
\intit_{t_0}^{y^0}\dd y_1\cdots\dd y_a=\intit_{t_0}^t\dd y_1\cdots\dd y_a 
-\int_{y^0}^t\dd y_1^0\int_{\R^{d}}\dd\vect{y_1}\intit_{t_0}^{y_1^0}\dd y_2\cdots\dd y_a
$$
Hence using this last identity and performing the change of variable $s\leftarrow s+1$, 
expression \refeq{premier.terme.monstre} leads to 
\begin{multline}\label{premier.terme.moins.monstre}
\sum_{\substack{r\ge 0;s\ge 1\\ r+s=N+1}}(-1)^{s}
\intit_{t_0}^{t}\dd y_1\cdots\dd y_r\intit_{t_0}^{t}\dd z_1\cdots\dd z_s 
\mathbb{V}(y_r\ldots y_1,x,z_1\ldots z_s)\\
-
\sum_{\substack{r\ge 1, s\ge 1\\ r+s=N+1}}(-1)^r
\int_{t_0}^{t}\dd z_1^0\int_{\R^{d}}\dd\vect{z_1} 
\int_{z_1^0}^t\dd y_1^0\int_{\R^{d}}\dd\vect{y_1} \\
\intit_{t_0}^{y_1^0}\dd y_2\cdots\dd y_r\intit_{t_0}^{z_1^0}\dd z_2\cdots\dd z_s\mathbb{V}(y_r\ldots y_1,x,z_1\ldots z_s)
\end{multline}
where $\mathbb{V}(y_r\ldots y_1,x,z_1\ldots z_s)$ denotes the product 
$$
\mathbb{V}(y_r\ldots y_1,x,z_1\ldots z_s):=
\phi_I^{p+1}(y_{r})\cdots\phi_I^{p+1}(y_{1})\phi_I(x)\phi_I^{p+1}(z_{1})\cdots\phi_I^{p+1}(z_{s})
$$

A similar computation (only replacing $y$ by $z$ and $r$ by $s$) shows that the second term of the right hand side of 
\refeq{somme.de.nouveau.monstrueuse} writes (modulo a factor $-i \left(\frac{i}{p+1}\right)^N$): 
\begin{multline}\label{deuxieme.terme.monstre}
\sum_{\substack{r\ge 1;s\ge 0\\ r+s=N+1}}(-1)^{s}
\intit_{t_0}^{t}\dd y_1\cdots\dd y_r\intit_{t_0}^{t}\dd z_1\cdots\dd z_s 
\mathbb{V}(y_r\ldots y_1,x,z_1\ldots z_s)\\
-
\sum_{\substack{r\ge 1, s\ge 1\\ r+s=N+1}}(-1)^r
\int_{t_0}^{t}\dd y_1^0\int_{\R^{d}}\dd\vect{y_1} 
\int_{y_1^0}^t\dd z_1^0\int_{\R^{d}}\dd\vect{z_1} \\
\intit_{t_0}^{y_1^0}\dd y_2\cdots\dd y_r\intit_{t_0}^{z_1^0}\dd z_2\cdots\dd z_s\mathbb{V}(y_r\ldots y_1,x,z_1\ldots z_s)
\end{multline}

Hence injecting \refeq{premier.terme.moins.monstre} and \refeq{deuxieme.terme.monstre} in \refeq{somme.de.nouveau.monstrueuse}, 
we finally find out that operator $\left(\frac{p+1}{i}\right)^{N+1}\varphi_{N+1}$ is given by 
\begin{multline}\label{presque.fin.monstre}
2 \sum_{\substack{r\ge 1;s\ge 1\\ r+s=N+1}}(-1)^{s}
\intit_{t_0}^{t}\dd y_1\cdots\dd y_r\intit_{t_0}^{t}\dd z_1\cdots\dd z_s 
\mathbb{V}(y_r\ldots y_1,x,z_1\ldots z_s)\\
+(-1)^{N+1}\intit_{t_0}^t\dd y_1\cdots\dd y_{N+1}\mathbb{V}(y_{N+1}\ldots y_1,x)
+\intit_{t_0}^t\dd z_1\cdots\dd z_{N+1}\mathbb{V}(x,z_1\ldots z_{N+1}) +A
\end{multline}
where $A$ denotes the operator 
\begin{multline}\label{A.monstre}
-\sum_{\substack{r\ge 1;s\ge 1\\ r+s=N+1}}(-1)^{s} 
\int_{\R^{d}}\dd\vect{y_1}\int_{\R^{d}}\dd\vect{z_1} 
\left(\int_{t_0}^{t}\dd z_1^0\int_{z_1^0}^t\dd y_1^0
+
\int_{t_0}^{t}\dd y_1^0\int_{y_1^0}^t\dd z_1^0\right)\\
\intit_{t_0}^{y_1^0}\dd y_2\cdots\dd y_r\intit_{t_0}^{z_1^0}\dd z_2\cdots\dd z_s\mathbb{V}(y_r\ldots y_1,x,z_1\ldots z_s)
\end{multline}
But we have the following identity 
$$
\int_{t_0}^{t}\dd z_1^0\int_{z_1^0}^t\dd y_1^0+\int_{t_0}^{t}\dd y_1^0\int_{y_1^0}^t\dd z_1^0
=
\int_{t_0}^{t}\dd z_1^0\int_{t_0}^t\dd y_1^0
$$
So inserting this last identity in the expression \refeq{A.monstre} of $A$ we get 
$$
A=-\sum_{\substack{r\ge 1;s\ge 1\\ r+s=N+1}}(-1)^{s}
\intit_{t_0}^{t}\dd y_1\cdots\dd y_r\intit_{t_0}^{t}\dd z_1\cdots\dd z_s 
\mathbb{V}(y_r\ldots y_1,x,z_1\ldots z_s)
$$
hence we finally see that \refeq{presque.fin.monstre} leads to 
$$
\left(\frac{p+1}{i}\right)^{N+1}\varphi_{N+1}=
\sum_{\substack{r\ge 0;s\ge 0\\ r+s=N+1}}(-1)^{s}
\intit_{t_0}^{t}\dd y_1\cdots\dd y_r\intit_{t_0}^{t}\dd z_1\cdots\dd z_s 
\mathbb{V}(y_r\ldots y_1,x,z_1\ldots z_s)
$$
which is exactly \refeq{ce.qu.on.veux.Heisenberg} at order $N+1$. 
\end{proof}

\section*{Acknowledgements} 
The author is very grateful to Sandrine Anthoine for careful reading of the manuscript and Fr\'ed\'eric H\'elein for helpful 
remarks and suggestions. 

\bibliographystyle{acm}

\end{document}